\documentclass[aps,pre,twocolumn,lengthcheck]{revtex4}
\usepackage{array}
\usepackage{amsmath}
\usepackage{graphicx}
\newcommand{\be}{\begin{equation}}
\newcommand{\ee}{\end{equation}}
\newcommand{\bea}{\begin{eqnarray}}
\newcommand{\eea}{\end{eqnarray}}
\newcommand{\f}{\frac}
\newcommand{\la}{\alpha}
\newcommand{\La}{\lambda}

\newcommand{\bra}{\langle}
\newcommand{\ket}{\rangle}

\newcommand{\ov}{\overline}

\newcommand{\sig}{\sigma}

\newcommand{\sa}{A}
\renewcommand{\sb}{B}
\newcommand{\tr}{\mathop{\rm tr}}

\begin{document}\author{A. C. Bertuola, J. X. de Carvalho, M. S. Hussein, 
M. P. Pato and  A. J. Sargeant}
\affiliation{Instituto de F\'isica, Universidade de S\~ao Paulo,
Caixa Postal 66318, 05315-970 S\~ao Paulo, S.P., Brazil}
\title{Level density for deformations of the Gaussian orthogonal ensemble}
\date{December  17, 2004}
\begin{abstract}
Formulas are derived for the average level density of deformed, or transition, 
Gaussian orthogonal random matrix ensembles. After some general
considerations about Gaussian ensembles we derive formulas for the
average level density for (i) the transition from the Gaussian orthogonal
ensemble (GOE) to the Poisson ensemble and (ii) the transition from the GOE to 
$m$ GOEs.
\end{abstract}
\maketitle

\section{Introduction}
Deformed random matrix ensembles were introduced by Rosenzweig and Porter 
\cite{Rosenzweig:1960,Porter:1965}
to classify the conservation of electronic spin
and orbital angular momentum in the spectra of complex atoms. 
Many varieties of deformed ensembles have since been constructed.
Recent reviews of deformed ensembles (they are also called transition
ensembles) can be found in Refs.~\cite{Kota:2001,Guhr:1998}. Earlier
results on deformed ensembles were reviewed in Ref.~\cite{Brody:1981}.
Amongst the applications they have found we mention the breaking of time
reversal invariance \cite{Hussein:1998} and the breaking
of symmetries \cite{Hussein:2000}.

In this paper we are concerned with the level density for deformed
Gaussian orthogonal random matrix ensembles. Although the level
density of the standard random matrix ensembles is not directly
related to the physical many particle level density it is essential 
to the proper unfolding of fluctuation measures \cite{French:1988}.
Unfolding is a transformation which leaves sequences of energy levels with
a constant average density. Normally, 
experimental energy levels {\em and} the energy levels obtained from random
matrix theory are unfolded (which removes the secular energy dependences) 
before fluctuation measures for both are calculated and compared.  
While the level densities of most random matrix ensembles are different,
the behaviour of 
the fluctuation measures of a wide class 
of ensembles is the same and hence denominated
universal. For instance, each of the non-Gaussian ensembles studied in
Ref.~\cite{Ghosh:2003} is shown to have a characteristic average level
density, however, after unfolding, the nearest neighbour spacing 
distributions and number variances of all are found to be identical 
to those of the
standard Gaussian random matrix ensembles \cite{Bohigas:1991}.

In the following we derive formulas for the level density of  
two deformations of the Gaussian orthogonal ensemble (GOE). The first 
describes a transition from the GOE to an ensemble with Poisson
fluctuation statistics, that is, 
from the GOE to an ensemble of diagonal matrices whose elements are 
independently Gaussian distributed \cite{Guhr:1989,Pato:1994}. Such random
matrix models are of interest to the analysis of experimental data 
\cite{Abd:1998} and of dynamical models \cite{Matsuo:1997} whose
fluctuation properties are intermediate between Poisson and GOE.
The second deformation describes the transition from the GOE to $m$ GOEs, 
that is, from the GOE to a block diagonal matrix with $m$ blocks each
of which is a GOE \cite{Guhr:1990,Hussein:1993,Hussein:1993b}. If the 
blocks are labelled by quantum numbers such as
angular momentum \cite{Rosenzweig:1960} or isospin \cite{Aberg:2004}
then the transition ensemble classifies their conservation 
or non-conservation. 
The $m$ GOE to GOE transition is also relevant to the analysis of symmetry
breaking in quartz blocks \cite{Ellegard:1996,Abd:2002,Schaadt:2003}.

Our method extends that of Wigner \cite{Wigner:1955}
who showed, by assuming that terms containing 
patterns of unlinked binary associations dominate the averages of 
the traces of powers of 
matrices, that the level density for certain random matrix ensembles
could be simply expressed as a Fourier transform
(see also Sec. III D Ref.~\cite{Brody:1981}). 
Previous results for the level density of deformed
GOEs have been derived using Stieltjes
transform methods \cite{Pastur:1972,Pandey:1981}. These methods are in
fact general enough to treat deformations of, and interpolations between,
any class of matrix ensemble. However, the formulas obtained in the
present paper have the advantage that that they are explicit and simple to 
evaluate numerically. Other methods for calculating the 
level density \cite{Guhr:1997,Kunz:1998} are restricted to
deformations of the Gaussian unitary ensemble (GUE).

\section{Interpolating Gaussian ensembles}
The joint probability distribution of the matrix elements of a matrix, $H$, 
for an interpolating Gaussian ensemble may be expressed as
\begin{equation}
P\left( H,\sa ,\sb \right) =Z^{-1}\left( \sa ,\sb \right) \exp
\left[ -\left( \sa \tr H^{2}+\sb \tr
H_{1}^{2}\right) \right],
\label{1}
\end{equation}
where $Z$ is a normalization factor and $\tr H$ denotes the trace of $H$. 
The structure of the matrix 
$H_{1}$ is chosen in such a way that it defines a subspace of $H$. 
The parameters $\sa $ and $\sb $ define the (energy) scale and degree 
of deformation. When $\sb \rightarrow 0$
the joint distribution becomes
\begin{equation}
P\left( H,\sa ,0\right) =Z^{-1}\left(\sa,0\right) \exp \left(
-\sa \tr H^{2}\right).
\label{7}
\end{equation}
If we assume that $H$ is real symmetric then the 
variances are given by
\begin{equation}
\ov{ H_{ij}^{2}} =\frac{1+\delta _{ij}}{4\sa },
\label{8}
\end{equation}
so that the limit $B\rightarrow 0$ defines the GOE.
When $\sb \rightarrow \infty $, the elements of $H_{1}$ vanish and $H$ is
projected onto a sparse matrix, $H_{0}$, whose elements are on the
complementary subspace of $H_{1}$. Therefore, the random matrices generated
by Eq. (\ref{1}) are the sum of two terms
\begin{equation}
H=H_{0}+H_{1},  \label{9}
\end{equation}
with the variances of the matrix elements of $H_{0}$ given by 
the r.h.s. of Eq.~(\ref{8}) and those of $H_{1}$ given by 
\begin{equation}
\ov{ H_{1ij}^{2}} =\frac{1+\delta _{ij}}{4\left( \sa
+\sb \right) }.  \label{10}
\end{equation}
This shows that when $\sb $ goes from zero to infinity, the ensemble
undergoes a transition from GOE to the Gaussian ensemble of sparse matrices
defined by the choice of the structure of $H_{1}$ or, equivalently, of 
$H_{0}$. It is also instructive to introduce the parameter
\be
\la=(1+\sb/\sa)^{-1/2}
\label{la}
\ee
which measures the relative strength of $H_1$ and $H_0$. The transition
from $\la$=0 to $\la$=1 corresponds to the transition from $\sb$=$\infty$
to $\sb$=0.
\section{Level Density for the GOE}
We now proceed to give a derivation of the semicircle law, valid for the GOE, 
which is equivalent to Wigner's for the ensemble of random
sign symmetric matrices \cite{Wigner:1955}.
In general, the average level density (ALD) may
be written as a Fourier transform,
\be\label{fourier}
\rho(E)=\f{1}{2\pi}\int_{-\infty}^{\infty}dkF(k)e^{-ikE},
\ee
where
\bea\label{Fk1}
F(k)&=&\int_{-\infty}^{\infty}dE\rho(E)e^{ikE}
\\&=&\sum_{n=0}^\infty\f{(ik)^n}{n!}
\int_{-\infty}^{\infty}dE\rho(E)E^n.
\label{Fk2}
\eea
Let $E_k$ denote the eigenvalues of an $N\times N$ matrix $H$, which
satisfies Eqs.~(\ref{7}) and (\ref{8}). 
From the exact expression, 
\be
\rho \left( E\right) =\f{1}{N}\sum_{k=1}^{N} \ov{\delta \left(E-E_{k}\right)}, 
\ee 
for the ALD, one obtains the following connection between
the moments of the eigenvalue distribution and the moments of the matrix
elements:
\begin{equation}
\ov{E^n}\equiv \int_{-\infty}^\infty dE\rho \left( E\right) E^{n}=\f{1}{N}
\ov{\tr H^{n}}.  \label{20}
\end{equation}
Substituting  Eq.~(\ref{20}) into Eq.~(\ref{Fk2}) we obtain
\begin{equation}
F\left( k\right) =\f{1}{N}\sum_{n=0}^{\infty }\frac{\left( ik\right) ^{n}}{n!}%
\ov{\tr H^{n}}.  
\label{22}
\end{equation}
It is also useful to note that by differentiating Eq.~(\ref{Fk1})
the moments of the eigenvalue distribution can be expressed as
derivatives of the Fourier transform of the ALD
evaluated at zero, that is,
\bea\label{mom}
\ov{E^n}=i^{-n}F^{(n)}(0),
\label{momier}
\eea
where $F^{(n)}(0)=\left[d^n F/dk^n\right]_{k=0}$.

For large matrices, $N$ $\gg 1$, the average of the trace is dominated by
the terms in which matrix indices may be contracted in such a way that we are
left with $s=\frac{n}{2}$ pairs of matrix elements. This means that we can
write
\begin{equation}
\ov{\tr H^n}
\simeq C_{s}N^{s+1}\sigma ^{2s},  \label{22a}
\end{equation}
where we have introduced the notation
\bea
\sig^2=\f{1}{4\sa}
\eea
for the variance of the off-diagonal matrix elements.
The factor $C_{s}$ counts the number of contractions that leads to pairs. We
observe that in a contraction $s-1$ indices are eliminated and $s+1$ remain
which explains the power of $N$ in the above expression and 
allows us to write
\begin{equation}
C_{s}=\frac{1}{s}\frac{\left( 2s\right) !}{\left( s-1\right) !\left(
s+1\right) !},  \label{23}
\end{equation}
where the binomial factor counts the number of ways $s+1$ indices can be
extracted out of $2s$ ones. The factor is then divided by $s$, that is, the
number of ways $s-1$ indices can be eliminated without changing the
contraction. 

Substituting Eq.~(\ref{22a}) in Eq.~(\ref{22}), we find that the
Fourier transform of the average density is given by
\bea\nonumber
F_1\left(a,k\right) &\simeq& 
\f{1}{ka/2}\sum_{s=0}^\infty\f{(-)^s(ka/2)^{2s+1}}{s!(s+1)!}
\\&=&\frac{1}{ka/2}J_{1}\left(ka\right),   
\label{F1}
\eea
where $J_{1}\left( x\right) $ is a first-order Bessel function and
\be
a=\sqrt{N/\sa}.
\label{a}
\ee
Using the formula 
\be\label{bate}
(1-x^2)^{1/2}=\f{1}{2}\int_{-\infty}^{\infty}y^{-1}J_1(y)e^{-ixy}dy,
\ee
the Fourier transform in Eq.~(\ref{fourier}) may evaluated to obtain 
the semi-circle law,
\be
\rho_1\left( a,E\right) =\left\{ 
\begin{array}{cl}
\frac{1}{\pi a^2/2}\sqrt{a^{2}-E^{2}},&\left|E\right| \le a \\ 
0, &\left| E\right| > a
\end{array}
\right., 
\label{wig}
\ee
where the radius of the semi-circle is given by Eq.~(\ref{a}).
The cumulative level density (CLD), $x(E)$, is defined by
\be
x(E)=\int_0^E\rho(E')dE',
\label{cum}
\ee
which for the semi-circle law is found to be
\be
x_1(a,E)=\left\{ 
\begin{array}{cl}
-1/2,& E < a\\
\f{E\sqrt{a^2-E^2}+a^2\arcsin(E/a)}{\pi a^2}.
&\left|E\right| \le a \\ 
1/2,&  E  > a
\end{array}
\right., \label{cum1}
\ee
The average number of levels up to energy $E$ is given in terms of the 
CLD by
$N[x(E)-x(-\infty)]$.
\section{The transition GOE to Poisson}
\begin{figure}
\includegraphics[width=.48\textwidth]{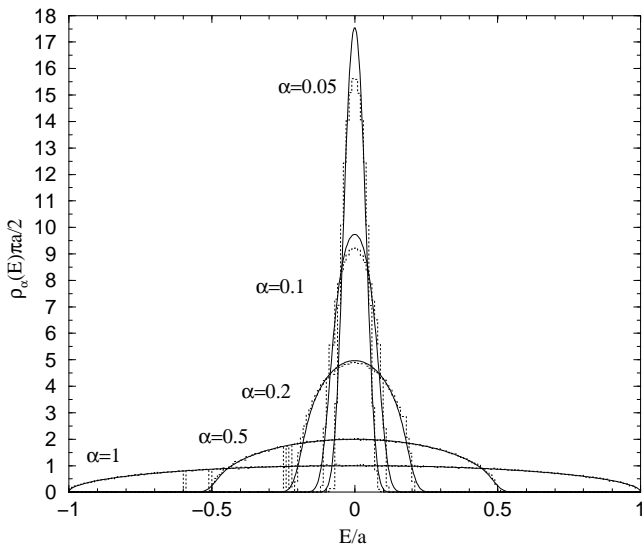}
\caption{Graph of $\rho_\la(E)$ for several values of $\la$ 
with $N$=1000. 
The calculations where performed with $A$=$N/4$ for which the radius
of the Wigner semi-circle is $a$=2. The solid lines were calculated using
Eq.~(\ref{poigoe}) and the dotted histograms by numerically diagonalizing 
an ensemble of 100 matrices.
\label{fig1}}
\end{figure}
Before considering the transition from Poisson to GOE we derive a
formula for the ALD for a more general case.
\subsection{Level density for $H_0+H_1$}
The trace of the $n$th power of Eq.~(\ref{9}) can be written as
\bea\nonumber
\tr
H^{n}&=& \tr\left(H_{0}+H_{1}\right)^{n}
\\&=&\sum_{l=0}^{n}\frac{n!}{l!\left( n-l\right) !}
\tr H_{0}^{l}H_{1}^{n-l}.
\label{56}
\eea
If  $H_{0}$ and $H_{1}$ and are statistically
independent we can write
\bea
\ov{\tr H_{0}^{l}H_{1}^{n-l}}
&=& \sum_j \ov{E_{0j}^l}\hspace{1mm}\ov{\bra j|H_{1}^{n-l}|j\ket} 
\label{t1}
\\&=& \ov{E_0^l}\hspace{1mm}\ov{\tr H_{1}^{n-l}},
\label{t2}
\eea
where $E_{0j}$ and $|j\ket$ are defined by the eigenvalue equation
$H_0|j\ket=E_{0j}|j\ket$ and $\ov{E_0^l}$ denotes the $l$th moment of the
eigenvalue distribution of $H_0$. Eqs.~(\ref{56}) and (\ref{t2})
allow us to write [cf. Eq.~(\ref{20})]
\bea
\nonumber
\f{1}{N}\ov{\tr H^{n}}
\approx\int_{-\infty }^{\infty }dE_{0}\rho_0(E_0)
\int_{-\infty }^{\infty }dE_{1}
\rho_1(E_{1}) \left(E_0+E_1\right)^n,
\\\label{60}
\eea
where $\rho_0$ and $\rho_1$ are the 
average level densities corresponding to
$H_0$ and $H_1$ respectively. Eq.~(\ref{60}) is only approximate because
in resumming the binomial series we have kept linked as
well as unlinked binary associations of $H_0$ and $H_1$. 
In the following subsection we show that for the case of
the transition GOE-Poisson the resulting discrepancy in the ALD is small.
We also show how to correct for this discrepancy. 

Substituting (\ref{60}) into (\ref{22}), we find
\bea\nonumber
F\left( k\right) 
&=&\int_{-\infty }^{\infty }dE_{0}
\rho_0(E_0)
\int_{-\infty }^{\infty }dE_{1}
\rho_1(E_{1}) e^{ik(E_0+E_1)}
\\&=& F_0(k)F_1(k), 
\label{Fprod}
\eea
where $F_0$ and $F_1$ are the Fourier transforms of 
$\rho_0$ and $\rho_1$ respectively. The ALD is obtained
by  substituting Eq.~(\ref{Fprod}) into Eq.~(\ref{fourier}).
Another representation is obtained
by noting that since Eq.~(\ref{Fprod}) is a product of 
Fourier transforms, the ALD of $H$ is given by
the convolution of the average level densities of $H_0$ and $H_1$: 
\be
\rho(E)=\int_{-\infty }^{\infty }dE'
\rho_0(E')
\rho_1(E-E') 
\label{conv}
\ee
The only assumption required to derive Eqs.~(\ref{Fprod}) and (\ref{conv}) is 
that the matrix elements of $H_0$ and $H_1$ be statistically independent. 
\subsection{Level Density for the transition GOE to Poisson}
To specialize the results of the last subsection to the transition GOE-Poisson,
$H_{0}$ is chosen to be the diagonal
matrix $H_{0ij}=E_{0i}\delta _{ij}$ whose eigenvalues, $E_{0i}$, are
independent random variables with Gaussian distribution 
\be
\rho_0(E)=\sqrt{\f{\sa}{\pi}}e^{-\sa E^{2}}.
\label{denpoi}
\ee
The variance of $H_0$ is thus
\be
\ov{H_{0ij}^2}=\f{\delta_{ij}}{2A}=a^2\f{\delta_{ij}}{2N}.
\label{varH0}
\ee
We choose  $H_{1}$ to be a diagonal-less matrix whose matrix elements are 
independent Gaussian variables with zero mean and variances
\begin{equation}
\ov{ H_{1ij}^{2}} =\la^2\frac{1-\delta _{ij}}{4\sa}
 =\la^2a^2\frac{1-\delta _{ij}}{4N}. 
\label{48}
\end{equation}

The ALD for $H_0$ alone is given by Eq.~(\ref{denpoi})
and the CLD, Eq.~(\ref{cum}), by
\be
x_0(E)=\f{1}{2}\mbox{erf}(\sqrt{A}E).
\label{cum0}
\ee
The Fourier transform of Eq.~(\ref{denpoi}) is 
\be
F_0(k)=e^{-k^2/4A}.
\label{F0}
\ee
Note that for fixed $a$, $A\to\infty$ as $N\to\infty$, so that
\be\label{rho0lim}
\rho_0(E)\xrightarrow[N\to\infty]{}\delta(E).
\ee

The ALD for $H_1$ alone is given by $\rho_1(\la a,E)$, 
that is, by the semicircle law [see Eq.~(\ref{wig})] with the radius modified  
($a\rightarrow \la a$) in accordance with Eq.~(\ref{48}) for the variance
of $H_1$ [cf. Eq.~(\ref{8})]. It's Fourier transform is $F_1(\la a,k)$ 
[see Eq.~(\ref{F1})].  The ALD for $H_1$
is given by the semicircle law in spite of the missing diagonal
because off-diagonal matrix elements dominate (for $N$ large enough).

\begin{figure}
\includegraphics[width=.48\textwidth]{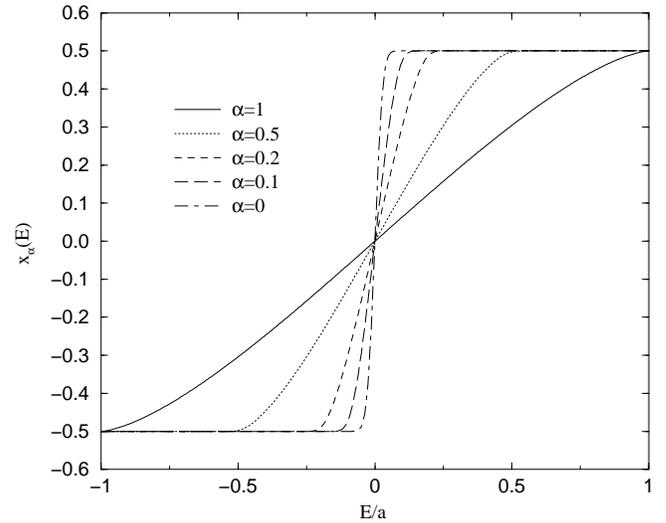}
\caption{Graph of $x_\la(E)$ calculated using Eq.~(\ref{cumft}) for
several values of $\la$ with $N$=1000. 
The calculations where performed with $A$=$N/4$ for which the radius
of the Wigner semi-circle is $a$=2. 
\label{fig2}}
\end{figure}

Using Eqs.~(\ref{Fprod}) and (\ref{fourier}) the ALD which
interpolates between the Gaussian and the semicircle as $\la$ varies between
0 and 1 is then found to be
\bea\label{pg}
\rho_\la\left( E\right)&=&\f{1}{2\pi}\int_{-\infty}^\infty dk
F_0(k)F_1(\la a,k)e^{-ikE}
\\\nonumber&=&
\f{2}{\pi}\f{\sqrt{\sa}}{\La}\int_{0}^{\infty }\f{dx}{x}
e^{-x^2/4\La^2}J_{1}(x) \cos\f{\sqrt{\sa} x E}{\La},
\\\label{poigoe}
\eea
where we have changed the integration variable to $x=\la ka$
and introduced the parameter
\be
\La=\la\sqrt{N}
\label{La}.
\ee 
In Fig.~\ref{fig1} we compare Eq.~(\ref{poigoe}) 
for $\rho_\la\left( E\right)$
with histograms constructed by numerically diagonalizing random matrices
for several values of $\la$ (see figure caption for details). 
It is seen that good agreement
is obtained, there being however a small discrepancy in the transition
region which is especially apparent in the graph for $\la$=0.05.
The CLD, Eq.~(\ref{cum}), may be expressed 
using Eq.~(\ref{poigoe}) as
\bea
x_\la(E)=
\f{2}{\pi}\int_{0}^{\infty }\f{dx}{x^2}
e^{-x^2/4\La^2}J_{1}(x) \sin\f{\sqrt{\sa} x E}{\La}.
\label{cumft}
\eea
This equation provides a more accurate manner of unfolding fluctuation
measures than the polynomial unfolding used in 
Ref.~\cite{Sargeant:1999qk}.
Eq.~(\ref{cumft}) is plotted for several values of $\la$ in Fig. \ref{fig2}.

The ALD may be expressed alternatively using the convolution formula, 
Eq.~(\ref{conv}). In particular, 
when $N\to\infty$ for fixed $\la$ and $a$ we find from Eq.~(\ref{conv}) 
and Eq.~(\ref{rho0lim}) that
\be\label{rhoalflim}
\rho_\la(E)\xrightarrow[N\to\infty]{}
\rho_1(\la a,E).
\ee
For finite $N$ Eq.~(\ref{conv}) yields 
\be\label{cvl}
\rho_\la(E)=\f{2\sqrt N}{\pi^{3/2}a^3\la^2}\int_{-\la a+E}^{\la a+E}
e^{-\f{NE'^2}{a^2}}\sqrt{\la^2a^2-(E-E')^2}dE'.
\ee

It it clear from these explicit formulae that the transition parameter is
$\la^2N$ as was already argued in the original paper of 
Rosenzweig and Porter \cite{Rosenzweig:1960}. The model described in this
section can be cast in the form 
\be
H=H_0+\f{\La}{N^c}H_1,
\label{c}
\ee
with $c=1/2$, by modifying the definitions of the variances, 
Eqs.~(\ref{48}). Ref.~\cite{Kunz:1998} considered the statistics of ensembles
of the form of Eq.~(\ref{c}) for arbitrary $c$.
In particular the asymptotic behavior of the ALD as $N\to\infty$ 
depends critically on $c$, Eq.~(\ref{rhoalflim}) being valid only 
for the special case $c=1/2$.

\begin{figure}
\includegraphics[width=.48\textwidth]{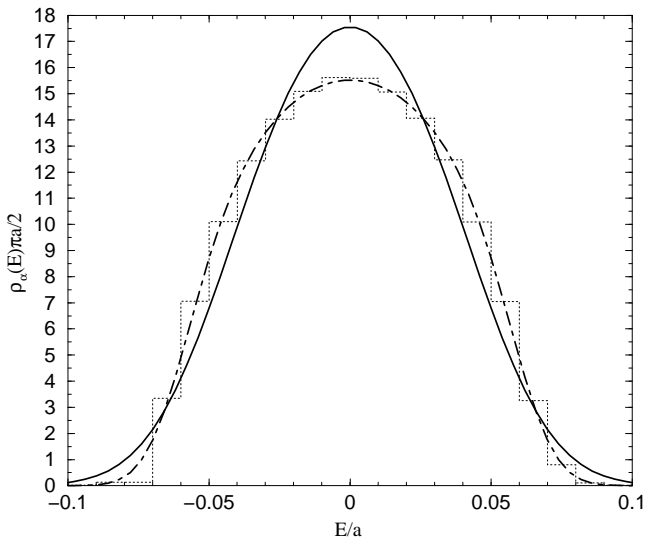}
\caption{Graph of $\rho_\la(E)$ showing the improvement obtained by 
demanding that the second and fourth moments of the eigenvalue distribution 
be exact for $\la$=0.05 and $N$=1000. The solid lines result from using
Eq.~(\ref{poigoe}) and the dot-dashed lines from
using the same equation modified in accordance with
Eqs.~(\ref{A'}) and (\ref{La'}). 
The dotted histograms were obtained by numerically diagonalizing 
an ensemble of 100 matrices.
\label{fig3}}
\end{figure}
Eq.~(\ref{poigoe}) may be improved by comparing the resulting lowest moments 
of the eigenvalue distribution with the average trace of the corresponding 
powers of the Hamiltonian. Considering the latter first we have
\bea\nonumber
\ov{\tr H^2}&=&\ov{\tr H_0^2}+\ov{\tr H_1^2}
\\\nonumber&=&N\ov{E_0^2}+C_1N^2\f{\la^2a^2}{4N}
\\&=&\f{N}{4A}\left(\La^2+2\right),
\label{H2}
\eea
and
\bea\nonumber
\ov{ \tr H^4}&=&\ov{\tr H_0^4}+4\ov{\tr{H_0^2H_1^2}}+\ov{\tr H_1^4}
\\\nonumber
&=&N\ov{E_0^4}+4\ov{E_0^2}C_1N^2\f{\la^2a^2}{4N}+C_2N^3\f{\la^4a^4}{16N^2}
\\&=&\f{N}{8A^2}\left(\La^4+4\La^2+6\right).
\label{H4}
\eea
Inserting Eq.~(\ref{poigoe}) into Eq.~(\ref{mom}) for the moments of the 
eigenvalue distribution we find that
\bea
\ov{ E^2}&=&\f{1}{4A'}\left(\La'^2+2\right),
\\\ov{ E^4}&=&\f{1}{8A'^2}\left(\La'^4+6\La'^2+6\right).
\label{E4}
\eea
Note that the coefficient of $\La'^2$ in Eq.~(\ref{E4}) is 6. This is 
a result of the use of Eq.~(\ref{60}) which included
linked binary associations of $H_0$ and $H_1$. Eq.~(\ref{20}) demands
that 
\bea
\ov{E^2}=\f{1}{N}\ov{\tr H^2},
\label{2nd}
\\\ov{E^4}=\f{1}{N}\ov{\tr H^4}.
\label{4th}
\eea
Solving Eqs.~(\ref{2nd}) and(\ref{4th}) for $A'$ and $\La'$ we see that 
Eq.~(\ref{poigoe}) will give the 2nd and 4th moments of the eigenvalues
distribution correctly if we make the substitution.
\bea\label{A'}
\f{1}{A}&\rightarrow& \f{1}{A'}=\f{1}{A}\left(1
+\La^2\left[1-\sqrt{1+\f{4}{\La^2}}\right]\right),
\\\La^2&\rightarrow&\La'^2=\La^2\f{\sqrt{1+\f{4}{\La^2}}}{
1+\La^2\left[1-\sqrt{1+\f{4}{\La^2}}\right]}.
\label{La'}\eea
\begin{figure}
\includegraphics[width=.48\textwidth]{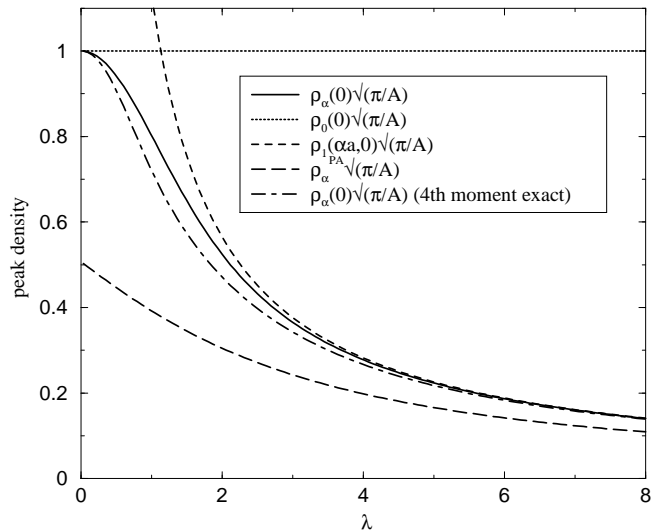}
\caption{Graph of the peak values of the density $\rho_\la(0)$, 
Eq.~(\ref{poigoe}), 
as a function of the transition parameter $\La$, Eq.~(\ref{La}).
For comparison we also show the limiting cases $\rho_0(0)$, Eq.~(\ref{denpoi}),
and $\rho_1(\la a,0)$, Eq.~(\ref{rhoalflim}),
as well as an interpolation formula
given by Persson and \AA berg \cite{Persson:1995} 
$\rho_{\la}^{\rm{PA}}$, Eq~(\ref{abden}). Finally, we show $\rho_\la(0)$ 
calculated using Eq.~(\ref{poigoe}) modified in accordance with
Eqs.~(\ref{A'}) and (\ref{La'}).
\label{fig4}}
\end{figure}
In Fig.~\ref{fig3} we show the improvement to the ALD obtained by using
Eqs.~(\ref{A'}) and (\ref{La'}) in  Eq.~(\ref{poigoe}) for the worst case of
Fig. \ref{fig1} ($\la$=0.05). We see that the modified
formula gives the ALD essentially exactly.

In Fig.~\ref{fig4} we show the peak value of the density of
states as a function of $\La$ (see figure caption). 
We see that Eq.~(\ref{poigoe}) in both its modified and unmodified versions
agree at the $\La$=0 limit. We also see that both versions of 
Eq.~(\ref{poigoe}) reach the $N\to\infty$ limit [Eq.~(\ref{rhoalflim})] at
roughly the same value of $\La$ ($\La$$\approx$6). However the two versions
deviate from each other in the transition region, the largest difference 
occuring around $\La$$\approx$1.5 ($\la$$\approx$0.05 for $N$=1000).
Also shown for
comparison is the interpolation formula for the ALD of Persson
and \AA berg \cite{Persson:1995}:
\be
\rho_{\la}^{\rm{PA}}
=\f{N^{\f{1}{2}}}{4\la N^{\f{1}{2}}+7N^{-1.5\la}}.
\label{abden}
\ee

\section{The transition GOE to $m$ GOEs}
\begin{figure}
\includegraphics[width=.48\textwidth]{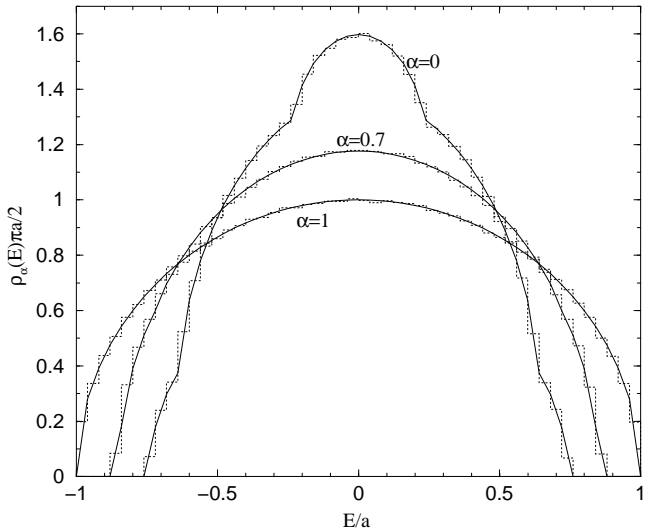}
\caption{
Graph of $\rho_\la(E)$ illustrating the transition 3 GOEs to
GOE. The three block have size $M_1$=50, $M_2$=400 and $M_3$=550.
The densities for $\la$= 0, 0.7, 1.0 are shown.
The calculations where performed with $A$=$N/4$ for which the radius
of the Wigner semi-circle is $a$=2. The solid lines were calculated using
Eq.~(\ref{40}) and the dotted histograms by numerical diagonalizing an ensemble
of 100 matrices.
\label{fig5}}
\end{figure}
We now calculate the ALD for the transition from the GOE to a superposition of 
$m$ GOEs. To proceed we again consider an ensemble of matrices of the 
form of  Eq.~(\ref{9}). Now $H_{0}$ is a block diagonal matrix consisting of 
$m$ blocks whose dimensions are $M_{i}$, $i=1,2,...,m$, with 
$\sum\limits_{i=1}^{m}M_{i}=N$.  
The elements of $H_0$ have zero mean and variances given by Eq.~(\ref{8}). 
We define $H_1$ to be zero where $H_0$ is non-zero and elsewhere its elements
have zero mean and variances given by Eq.~(\ref{10}). 

To obtain a formula for the ALD we note that
\bea
\ov{\tr H^{2}} 
&=&\sum_{j,k=1}^{N}\ov{ H_{jk}^{2}}
\\&=&N^2\sum_{i=1}^{m}\f{M_{i}}{N}\sigma_i^2  \label{30}
\eea
with
\begin{equation}
\sigma _{i}^{2}=
\f{1}{4\sa}\left[\f{M_{i}}{N}+\la^2(1-\f{M_{i}}{N})\right].  \label{44}
\end{equation}
Considering a single line of $H$ the $\sigma_i^2$ consist of a 
term which is the product of the variance of a single non-zero element of 
$H_0$ with the probability for being in block $i$ plus a term which 
is the product of the variance of a single non-zero element of $H_1$
with the probability for being outside block $i$. In the sum in Eq.~(\ref{30})
the $\sigma_i^2$ are weighted with the fraction of lines which find 
themselves in block $i$.

For the fourth power of $H$ we find
\bea
\ov{\tr H^{4}} 
&=&2\sum_{j,k,l=1}^{N}\ov{
H_{jk}^{2}} \ov{ H_{jl}^{2}}
\\\label{32a}&=&2\sum_{j=1}^{N}\left[ \sum_{k=1}^{N}\ov{ H_{jk}^{2}}
\right] ^{2}
\\&=&2N^3\sum_{i=1}^{m}\f{M_{i}}{N}\sigma_i^4.
\label{32}
\eea
Although the authors were unable to obtain an equation analogous
to Eq.~(\ref{32a}) for higher powers of $H$, Eqs.~(\ref{30}) and (\ref{32})
strongly suggest that   
\begin{equation}
\ov{\tr H^{2s}} 
\simeq
C_{s}N^{s+1}\sum_{i=1}^{m}\f{M_{i}}{N}\sigma_i^{2s}, \label{34}
\end{equation}
for large $N$.

Substituting Eq~(\ref{34}) into Eq.~(\ref{22}) and again using
Eqs.~(\ref{F1}), (\ref{fourier}) and (\ref{bate}) we obtain 
\begin{equation}
\rho\left( E\right) =\sum_{i=1}^{m}\f{M_i}{N}
\rho_1\left( a_{i},E\right), 
\label{40}
\end{equation}
where $\rho_1$ is given by Eq.~(\ref{wig}) and
\bea
a_{i}&=&a^2\left[\f{M_i}{N}+\la^2(1-\f{M_i}{N})\right]
\\&=&\f{1}{A}\left[M_i+\La^2(1-\f{M_i}{N})\right].
\label{ai}
\eea
From Eq.~(\ref{40}) it follows that the CLD, Eq.~(\ref{cum}),
is given by
\be\label{cumla2}
x\left( E\right) =\sum_{i=1}^{m}\f{M_i}{N}
x_1\left( a_{i},E\right) ,
\ee
with $x_1$ given by Eq.~(\ref{cum1}).

In Fig.~\ref{fig5}, we compare Eq.~(\ref{40}) for the ALD
with numerical simulations for several values of the $\la$ (see figure
caption for details) and excellent agreement is obtained. 
\section{Conclusion}
In conclusion, by assuming that terms containing 
patterns of unlinked binary associations dominate the the averages of
traces of powers of 
matrices, we have derived formulas for the average level density for
two deformations of the Gaussian orthogonal ensemble. The 
first describes the transition from the Gaussian orthogonal
ensemble to the Poisson ensemble and the second the transition from the GOE to 
$m$ GOEs. The formula obtained are in excellent agreement with numerical
simulations.
\begin{acknowledgments}
This work was carried out with support from FAPESP, the CNPq and the
Instituto de Mil\^enio de Informa\c c\~ao Qu\^antica - MCT.
\end{acknowledgments}
\bibliography{rmt,compound,sargeant,books,sd}
\bibliographystyle{apsrev}








\end{document}